\documentclass[aps,prb,groupedaddress,twocolumn,amsmath,amssymb,eqsecnum,floatfix]{revtex4}
\usepackage{graphicx}
\begin{document}

%\preprint{}

%otv \title{Supersolid Behavior from Superfluidity in grain boundaries}
\title{Non-classical Rotational Inertia in a Two-dimensional Bosonic Solid Containing
Grain Boundaries}


\author{Chandan Dasgupta}
\email{cdgupta@physics.iisc.ernet.in}
%\thanks{}
\altaffiliation{Also at Condensed Matter Theory Unit, Jawaharlal Nehru Centre
for Advanced Scientific Research, Bangalore 560064, India}
\affiliation{Centre for Condensed Matter Theory, Department of Physics, 
Indian Institute  of Science, Bangalore 560012, India}
\author{Oriol T. Valls}
\email{otvalls@umn.edu}
\altaffiliation{Also at Minnesota Supercomputer Institute, University of Minnesota,
Minneapolis, Minnesota 55455}
\affiliation{School of Physics and Astronomy,
University of Minnesota, Minneapolis, Minnesota 55455}

\date{\today}

\begin{abstract}

We study the occurrence of 
non-classical rotational inertia (NCRI) arising from superfluidity 
along grain boundaries in a two-dimensional bosonic
system. We make 
use of a standard mapping between the zero-temperature properties of this 
system and the statistical mechanics of interacting 
vortex lines in the mixed phase of a type-II superconductor.
In the mapping, the liquid phase of the vortex system corresponds to
the superfluid bosonic phase. We consider numerically obtained
polycrystalline configurations of the vortex lines in which
the microcrystals are separated by liquid-like
grain boundary regions which widen as the
vortex system temperature increases. The NCRI of the corresponding zero-temperature 
bosonic systems can then be numerically evaluated by solving the equations of superfluid
hydrodynamics in the channels near the grain boundaries. We find that the NCRI increases very 
abruptly as the liquid regions in the vortex system (equivalently, superfluid regions
in the bosonic system) form a connected, system-spannig structure with one or more
closed loops. The implications of these results for experimentally
observed supersolid phenomena are discussed.

\end{abstract}

%\pacs{  }
%\keywords{}

\maketitle

\section{Introduction}
\label{intro}

The observation~\cite{chan1,chan2}
of non-classical rotational inertia (NCRI) in torsional 
oscillation experiments on solid $^4{\rm He}$  created a great deal of
interest~\cite{balibar} in the possibility of occurrence of a ``supersolid'' phase in which 
crystalline order and superfluidity coexist. %otv
Considerable subsequent work has 
sustained this interest. The occurrence of NCRI in
solid $^4{\rm He}$ has been confirmed in many experiments~\cite{reppy,kojima,kubota,kondo,hunt,beamish}
and the dependence of the measured NCRI fraction (NCRIF) on various factors, such as the
method of sample growth, its annealing, 
frequency and amplitude %otv too many of the samples
of the torsional oscillator and the amount of $^3{\rm He}$ impurities 
present, 
have been studied in great detail. However, despite these extensive studies, the 
microscopic origin of the observed NCRI signal remains quite controversial. In particular, it is
not clear whether the superfluid component in supersolid $^4{\rm He}$ is distributed
uniformly throughout the sample, or present only near structural defects such as dislocations
and grain boundaries. A suggestion that the observed phenomena
were related to an old theoretical proposal~\cite{lifshitz} of superfluidity 
arising from mobile zero-point vacancies in the crystal has been contradicted by 
results of quantum Monte Carlo simulations\cite{qmc1,qmc2} that show that the concentration
of vacancies is too low to account for the measured NCRIF. On the other hand, the observed
dependence~\cite{reppy} of the magnitude of the NCRIF on the quality
of the sample (samples with higher degree of crystalline order exhibit smaller NCRIF) argues 
in favor of a mechanism of superfluidity in which defects play a major role. An important
role for the defects is also indicated by the observation~\cite{beamish,beamish2} 
of a close correspondence between the
onset of NCRI and an increase of the shear modulus of the solid, and by an enhancement of both
these effects when the concentration of $^3{\rm He}$ impurities is increased. It has been 
suggested~\cite{beamish2} that
both the occurrence of NCRI and the simultaneous increase in the shear modulus arise from a stiffening
of a network of dislocation lines, and that this stiffening is assisted by
$^3{\rm He}$ impurities which pin the dislocation lines by binding to them.

The question of whether the experimentally observed supersolid
behavior in $^4{\rm He}$ can arise from the presence of structural defects has been investigated, 
on the theoretical side, in
quantum Monte Carlo studies~\cite{qmc3,qmc4,qmc5} that show that superfluidity can indeed occur
along grain boundaries and in the cores of dislocations. There have been a few 
attempts~\cite{gb1,gb2,shevchenko,toner1,toner2}
to calculate the macroscopic properties (such as the NCRIF) of a system in which superfluidity 
occurs along a network of grain boundaries or  dislocation lines. Since these structural defects
form disordered complex networks, a calculation of
the rotational inertia of a superfluid confined in a network of irregular-shaped channels
is necessary for understanding whether a defect-based mechanism can provide a consistent
explanation of the observed results. To our knowledge, no such calculation for realistic defect
network structures currently exists in the literature.

In this paper we study  
the possibility of occurrence of ``supersolid'' behavior, similar to that observed in solid 
$^4{\rm He}$,  arising
from superfluidity along grain boundaries in a two-dimensional bosonic system. 
We make use of a well-known mapping~\cite{nelson,nv,lv} between the statistical
mechanics of a system of interacting vortex lines in the mixed phase of type-II superconductors and
the zero-temperature quantum mechanics of a {\it two-dimensional} system of bosons. 
Specifically, we consider a highly anisotropic layered superconductor in the presence 
of a magnetic field perpendicular to the 
layers. The system then
consists of a collection of vortex lines oriented, on the average, normal to the layers. 
The statistical mechanics of these vortex lines can be mapped~\cite{nelson,nv,lv} onto
the quantum mechanics of a two-dimensional system of interacting bosons at zero temperature.
The mapping can also be used if a small concentration of columnar
pinning centers normal to the layers is present in the vortex system. In that case, the equivalent bosonic system
contains a small concentration of
randomly located point pinning centers that produce a random external potential for the bosons.
In this mapping, the liquid phase of the vortex system corresponds to the superfluid 
phase of the bosons.
The equilibrium properties of the vortex system, both with and without columnar pinning, 
have been investigated in a large number of
theoretical~\cite{nelson,nv,lv}, experimental~\cite{expt1,expt2,expt3} and 
numerical~\cite{dv1,dv2,dv3,sim1,sim2} studies. These results establish, as we will show
below, that the zero-temperature, two-dimensional, interacting bosonic system with random pinning centers
exhibits a polycrystalline state with superfluidity along grain boundaries
over a suitable range of system parameters.

Our earlier studies~\cite{dv1,dv2,dv3} of the 
vortex system in the presence of columnar pins
provide us with several realistic configurations
of the network of grain boundaries. We find that some of these polycrystalline states survive as
metastable states when the random pinning potential is turned off, as explained
below. This case maps then onto the impurity-free bosonic system,
which is similar to the experimentally studied $^4{\rm He}$ case. We then study the NCRIF arising from
the superfluid regions along the grain boundaries by numerically solving the equations of superfluid
hydrodynamics~\cite{fetter,hydro} in the geometry specified by the superfluid channels in the sample. This
allows us to determine the NCRIF as a function of system parameters. 
At low temperatures (meaning the temperature of the vortex system, {\it not} 
that of the
two-dimensional bosonic system), the liquid regions (superfluid regions in the equivalent 
two-dimensional system of bosons) near the grain boundaries are completely absent or small, and
the NCRIF is vanishingly small. As the temperature is increased towards the melting temperature of the 
vortex lattice in the absence of any pinning, a kind of ``premelting'' occurs near the
grain boundaries, so that the area covered by the liquid regions increases. This increases the
connectivity of the network of liquid channels and causes the NCRIF to 
increase as the channels open up throughout the entire sample. However,
the NCRIF remains vanishingly small as long as the liquid regions remain 
isolated from one another. Our main 
result is that the NCRIF exhibits a sharp jump to a measurable value of a few percent 
when the growing liquid regions
percolate across the system to form one or more closed channels of size 
comparable to the size of the system.
The behavior of the NCRIF as a function of the temperature of the vortex system is qualitatively similar
to that seen in the experiments on $^4{\rm He}$. Our work, thus, shows that supersolid behavior
similar to that observed in solid $^4{\rm He}$ can occur from superfluidity in grain boundaries 
in a two-dimensional system of bosons for an appropriate choice of parameters.

The rest of the paper is organized as follows. In section~\ref{mapmeth}, we provide the details of the
vortex-to-boson mapping used in our work, define the model we consider and describe the numerical methods
used in our calculations. The results of our study are described in detail in section~\ref{results}. 
Section~\ref{summary} contains a discussion of the implications of our results in the
context of current research on supersolidity in $^4{\rm He}$.
 
\section{Mapping, models and methods}
\label{mapmeth}

In this section, we first describe in detail the mapping between the statistical mechanics of
a collection of vortex lines and the zero-temperature quantum mechanics of a two-dimensional
system of bosons. We then discuss how existing results for the vortex system can be used to 
infer the occurrence of a ``supersolid'' phase with superfluidity along grain boundaries in the
bosonic system under suitable conditions. In our study, information about the behavior of the
bosonic system is obtained, through the vortex-boson mapping, from calculations carried out for
the vortex system. At the end of this section, we provide some details of the model used in our
study of the vortex system and the method of calculation we have used.

\subsection{Mappings and Models}

It is well established~\cite{nelson,nv,lv} that the partition function of a system of interacting vortex
lines can be written as that of a two-dimensional system of interacting bosons. We review
here only the necessary details. Let us consider a
system of $N$ vortex lines in a type-II superconductor in the mixed phase, with the magnetic field
in the $z$-direction. We also assume that $N_p$ columnar pinning centers oriented along 
the $z$ direction may be present in the sample
(this random pinning potential will be turned off in the numerical work described below). 
Let the two-dimensional vector ${\bf r}_j(z)$ denote the transverse position
of the $j$-th vortex line in the $xy$-plane at $z$ ($0 \leq z \leq L$ where $L$ is the thickness of the
sample in the $z$-direction). The free energy of the system of vortex lines has the form
\begin{eqnarray}
F_N&=&\frac{1}{2} \sum_{j=1}^N \epsilon
\int_0^L \left | \frac{d{\bf r}_j(z)}{dz}
\right |^2 dz \nonumber \\
&+& \frac{1}{2} \sum_{i\ne j} \int_0^L V(|{\bf r}_i(z)-{\bf r}_j(z)|) dz
\nonumber \\
&+&\sum_{j=1}^N \int_0^L V_d({\bf r}_j(z)) dz.
\label{fe}
\end{eqnarray}
Here, $\epsilon$ is the tilt modulus of the vortex lines, $V(r)$ is the 
interaction potential between two vortex lines separated by transverse distance $r$, and $V_d({\bf r})$
is a pinning potential produced by the columnar pinning centers (for columnar pins oriented in the
$z$-direction, the pinning potential does not depend on $z$). We assume periodic boundary condition in
the $z$-direction, which implies that the positions $\{{\bf r}_j(0)\}$ and $\{{\bf r}_j(L)\}$ at the two
ends of the sample must match modulo a permutation. The partition function of the system of vortex
lines at temperature $T$ is then given by
\begin{widetext}
\begin{eqnarray}
Z_v(T) = \frac{1}{N!} \sum_P \prod_{j=1}^N \int_{{\bf r}_j(L)={\bf r}_{P(j)}(0)}
{\mathcal{D}}\{{\bf r}_j(z)\} \exp[-F_N/T],
\label{partfn}
\end{eqnarray}
where the functional integrals are over all vortex-line configurations that satisfy the boundary conditions
${\bf r}_j(L)={\bf r}_{P(j)}(0)$ for all $j$, $P$ representing a permutation of the indices 
$1,2,\ldots,N$. 

To see the connection of this problem with the quantum mechanics of  bosons, let us consider a
two-dimensional system of $N$ identical bosonic particles of mass $m$, with pairwise interactions
given by the potential $V(r)$, in the presence of an external 
impurity potential $V_d({\bf r})$. In the
path integral representation, the partition function of this system at temperature $T_b$ can be
written as
\begin{eqnarray}
Z_b(T_b) &=& \frac{1}{N!} \sum_P  \prod_{j=1}^N \int_{{\bf r}_j(\Lambda)={\bf r}_{P(j)}(0)}
{\mathcal{D}}\{{\bf r}_j(\tau) \}
\exp \left[-\frac{1}{\hbar} \left\{ \sum_{j=1}^N \int_0^\Lambda \frac{m}{2} \left | 
\frac{d{\bf r}_j(\tau)}{d\tau} \right |^2 d \tau \right . \right .\nonumber \\
&+& \left . \left .\frac{1}{2} \sum_{i\ne j} \int_0^\Lambda V(|{\bf r}_i(\tau)-{\bf r}_j(\tau)|) d\tau
+\sum_{j=1}^N \int_0^\Lambda V_d({\bf r}_j(\tau)) d\tau \right \} \right],
\label{partfnb} 
\end{eqnarray}
where ${\bf r}_j(\tau)$ now represents the position of the $j$-th boson at ``imaginary time'' $\tau$
and $\Lambda= \hbar/T_b$. A comparison of this expression with that in Eq.~(\ref{partfn}) shows that 
the two are the same when one makes the identifications
\begin{equation}
\label{map} %otv
\epsilon \to m,\,\, T \to \hbar, \,\, L \to \Lambda = \hbar/T_b. 
\end{equation}
\end{widetext}
Thus, the thermodynamic limit, $L \to \infty$, in the vortex system corresponds to the zero
temperature ($T_b=0$) limit in the boson system. The tilt modulus of the vortex lines plays the role 
of the mass of the bosons and the temperature $T$ of the vortex system plays the role of $\hbar$.
The pairwise interaction $V(r)$ in the system of bosons is the same as that between two vortex lines and
the external potential $V_d({\bf r})$ is also the same in the two systems. The interaction between two
straight vortex lines parallel to each other is repulsive and its dependence of the transverse
separation is given by $V(r) \propto K_0(r/\lambda)$ where $K_0$ is a Bessel function and $\lambda$ 
the in-plane London penetration depth of the superconductor. Thus, the pair interaction potential 
in the corresponding boson problem is repulsive, logarithmic in $r$ for $r$ much smaller than
$\lambda$, and falls off exponentially as $\sim \exp(-r/\lambda)$ when $r$ is larger than $\lambda$.
Liquid phases in the vortex system correspond to %otv sentences added
superfluid bosonic phases, while vortex crystalline phases 
correspond to a bosonic crystal.

This mapping between the two systems allows us to draw certain conclusions about the behavior of
the two-dimensional bosonic system from the wealth of information available from existing
experimental~\cite{expt1,expt2,expt3} and numerical~\cite{dv1,dv2,dv3,sim1,sim2} studies of the
mixed phase of type-II layered superconductors (high-temperature cuprate superconductors in particular)
in the presence of random columnar pinning centers oriented parallel to the magnetic field
which is perpendicular to the layers.
These studies establish that when the concentration of pinning centers is small compared to 
that of vortex lines, the vortex system
exhibits a Bose glass (BoG) phase at low temperatures, and a vortex liquid phase at high temperatures. 
In the vortex-boson mapping, the
vortex liquid phase corresponds to the superfluid phase of the two-dimensional bosonic system. Both
experiments~\cite{expt1} and numerical studies~\cite{dv1,dv2,dv3} show that the BoG 
phase in the vortex system has a polycrystalline structure with grain boundaries separating crystalline
domains (see, for example, Fig.~2 of Ref.~\onlinecite{expt1} and Fig.~1 of Ref.~\onlinecite{dv1}). 
Experiments also show the existence of a ``vortex nanoliquid'' phase~\cite{expt2,expt3} 
near the boundary between the BoG and vortex liquid phases (see Fig.~2 of Ref.~\onlinecite{expt3}). 
From direct visualization of the flow of transport current in the system, it has been
established~\cite{expt3} that the ``vortex nanoliquid'' is 
characterized by the simultaneous presence of solid- and liquid-like regions in the system. 
This phase is distinguished from the homogeneous vortex liquid by the presence of interconnected
``droplets'' of vortex liquid caged inside a solid matrix formed by other vortices. The experiments
can not determine the positions of the droplets of vortex liquid relative to the grain boundaries that
are known to exist in the BoG phase. This information is provided by our earlier numerical
studies~\cite{dv1,dv2} of the vortex system. These studies reproduce all the experimentally 
observed features and in addition, show that the liquid
regions in the ``vortex nanoliquid'' phase lie along the grain boundaries (see, for example,
Fig.~7 of Ref.~\onlinecite{dv2}). Since the vortex liquid
corresponds to the superfluid in the bosonic system, these results establish that a zero-temperature
two-dimensional system of bosons, interacting via a repulsive pair potential proportional to $K_0(r)$
and in the presence of a small concentration of attractive pinning centers, exhibits a polycrystalline
phase with superfluidity along the grain boundaries for a suitable choice of system parameters.
In the present study, we examine whether this behavior persists in the bosonic system when the
pinning potential is turned off, and calculate how the rotational inertia of the system is affected
by the presence of droplets of superfluid lying along the grain boundaries.

\subsection{Methods}

Since our results for the bosonic system are obtained from studies of a system of vortices using
the methods of our earlier work~\cite{dv1,dv2,dv3}, we provide here a summary of the vortex model we
consider and the numerical method we use to study its equilibrium behavior. The details of both the
model and the method of calculation may be found in Ref.~\onlinecite{dv0}.

%otv some rewrite of below paragraph
The system studied consisted of a set 
of vortex lines in a highly anisotropic, layered superconductor with the magnetic
field perpendicular to the layers. The vortex lines are
formed by stacks of ``pancake'' vortices located
on the layers. The Josephson interaction between pancake vortices 
on different layers was neglected.
The structure and thermodynamic properties of the vortex system are 
determined from numerical
minimization of a model free energy functional 
of the Ramakrishnan-Yussouff form~\cite{ry} that 
expresses the free energy of the system as a 
functional of the time-averaged local density of the pancake vortices. 
For columnar pins perpendicular to the layers, the pinning potential is the
same on all the layers. This implies that the time-averaged density 
distribution is also the same
on all the layers. This makes the problem effectively two-dimensional~\cite{dv0}. 
Different phases
of the vortex system correspond to different local minima of the free energy 
and phase transitions 
are signalled by crossings of the free energies of different local minima. From the density
distribution at a local minimum representing a particular phase of the system, the
positions of the vortices in that phase 
are generated by locating the points at which the density exhibits local
peaks. This allows the study the real-space structures of different phases. The heights of the
local density peaks are used to determine whether the vortices in a region of the sample are in
a solid or liquid state.

Since our method of calculation is
designed for the vortex system, the parameter that we  control is the  %otv
temperature $T$ of the vortex system.
Changing this temperature is equivalent to ``changing the value of $\hbar$'' in the 
corresponding zero $T$ bosonic system via the mapping Eq.~(\ref{map}). %otv 
This should be interpreted as changing system parameters in such a way that the relative
importance of quantum effects is modified (``increasing $\hbar$'' implies increasing the 
importance of
quantum fluctuations).
%otv lines below canceled (?)
%%So, changing the temperature in the vortex system amounts to following a possibly 
%complicated path in the parameter
%space of the equivalent system of bosons. 
This correspondence should be kept in mind in the interpretation of our
results in the context of the bosonic problem.

\section{Results}
\label{results}

As mentioned above, all our results for the two-dimensional bosonic system have
been obtained from numerical calculations carried out for the equivalent vortex system.
In our earlier studies~\cite{dv1,dv2,dv3}, we considered a system of vortex lines in the
presence of randomly placed columnar pinning centers and found polycrystalline BoG 
states when the areal density of the pinning centers is much smaller than that of the 
vortex lines. Since quenched disorder arising from
the presence of pinning centers is not present in the $^4{\rm He}$ samples studied
experimentally in the context of supersolid behavior, we first investigated whether the
polycrystalline states found in our earlier studies survive when the random pinning
potential is turned off.

%otvr this paragraph edited
We use for this purpose two of the vortex configurations obtained and 
studied in Ref.~\onlinecite{dv3}. These correspond to results
%obtained 
for a relative concentration $c = 1/8$ of columnar pins 
($c = n_p/n_v$ where $n_p$, $n_v$ are, respectively, the areal densities
of pinning centers and vortex lines) in a system that
contained 4096 vortices (therefore 512 columnar pins). 
We consider, for our starting
point, vortex configurations at a vortex %otv
temperature $T=17.0 K$ at which
point (for $c=1/8$) the vortex system is well into the BoG phase,
see the phase diagram  in Fig.~1 of Ref.~\onlinecite{dv3} (to set the
temperature scale, we note that when no pinning is present, the vortex system considered  
undergoes a first-order melting transition from a crystalline Abrikosov lattice
to a vortex liquid at $T=18.4K$.)
%cdr Naively, one would expect that if the pinning potential is set to zero and the
%vortex system allowed to evolve towards a new free energy minimum,
%then it would converge to 
A perfectly crystalline state without grain boundaries is
% which would be 
the absolute minimum of the free energy in the absence of pinning. 
However % we have found that
if, using the polycrystalline BoG configuration as a starting point, one 
reduces the strength of the pinning potential 
to zero in sufficiently small steps while running at
every step the
free-energy minimization routine, one ends up, when zero pinning
strength is finally reached, with a {\it polycrystalline} sample, 
with well-defined grain
boundaries. This polycrystalline state is of course metastable, %minimization
%after simply setting the columnar pin potential to zero leads indeed
%to the Abrikosov lattice state mentioned above. Considerable effort
%had to be spent finding out, by trial and error, how small the steps
%by which the strength of the pinning potential is reduced must be in order
%for the system to remain ``stuck'' in the metastable polycrystalline
%state, as desired. This is difficult
but we find that as 
%cdr after such a metastable state is obtained 
it is warmed up (we use steps of 0.2 K),
% and that
the sample remains in the local polycrystalline free energy minimum
%cdr changes below
(within the finite accuracy of our numerical minimization procedure)
%corresponding to the polycrystalline state
until near $T=18.4K$ when, 
as mentioned above, the microcrystals melt. 
% and the samples become fully liquid. %These observations 
%about the local stability of the polycrystalline states are limited by the finite numerical
%accuracy of the procedure used to locate the free energy minima. 
%The
%iterative minimization procedure is stopped
%when the largest fractional change in the local density variables
%falls below\cite{dv3} a small tolerance factor. 
%We have checked that these states do not change when
%this tolerance factor is reduced, but it obviously cannot be brought down to zero. 
%Hence, some %we can not mathematically rule
%possibility remains that these polycrystalline states are not true
%local minima of the free energy. 
%This, however, is not a crucial issue here because 
Even if these polycrystalline states are not true local minima of the free energy, 
they can be made so by the introduction of a few
pinning centers that are always present in any physical sample.

\begin{figure}[tbh]
\includegraphics [width=3.0in,angle=0]{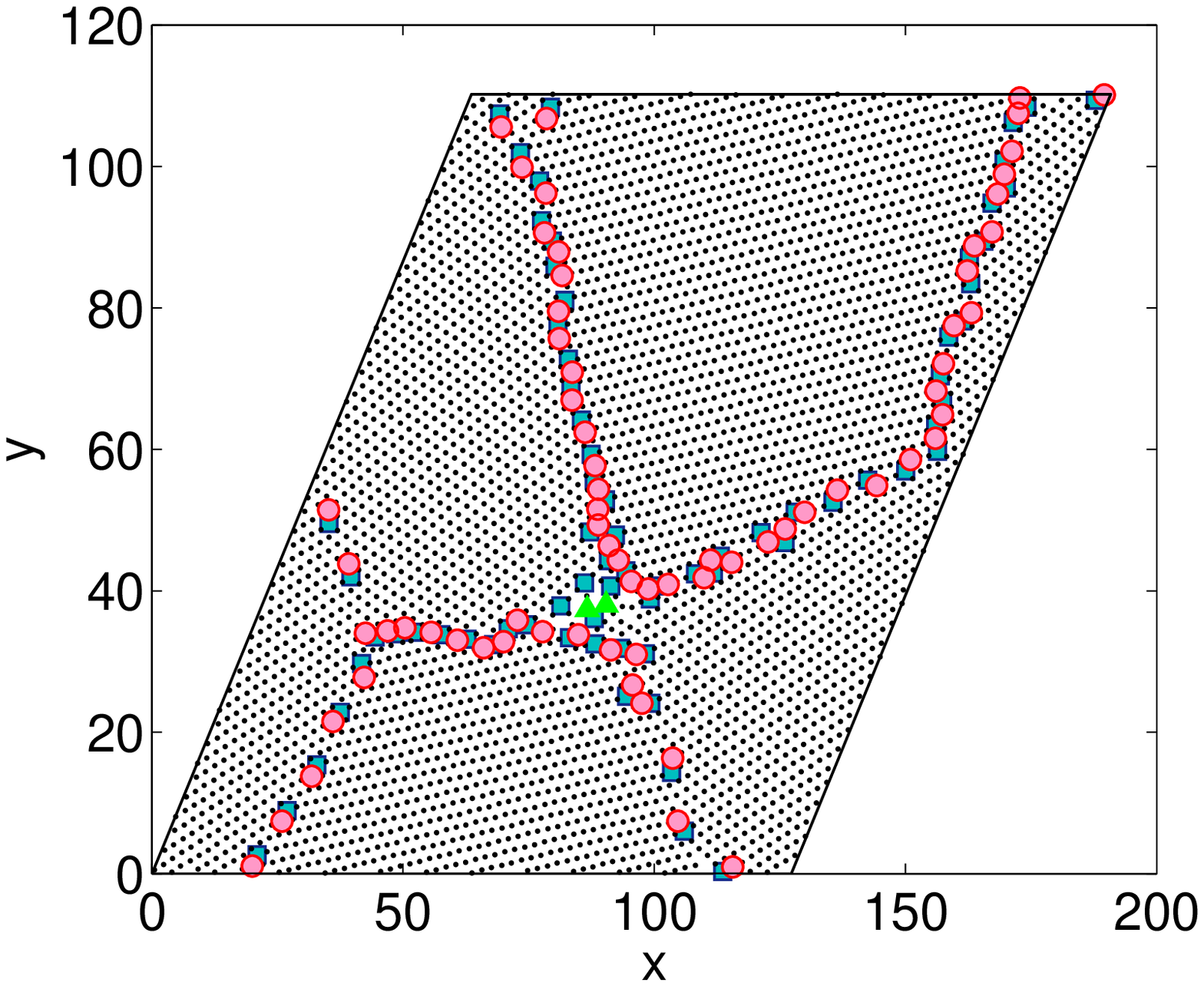}
\includegraphics [width=3.0in,angle=0] {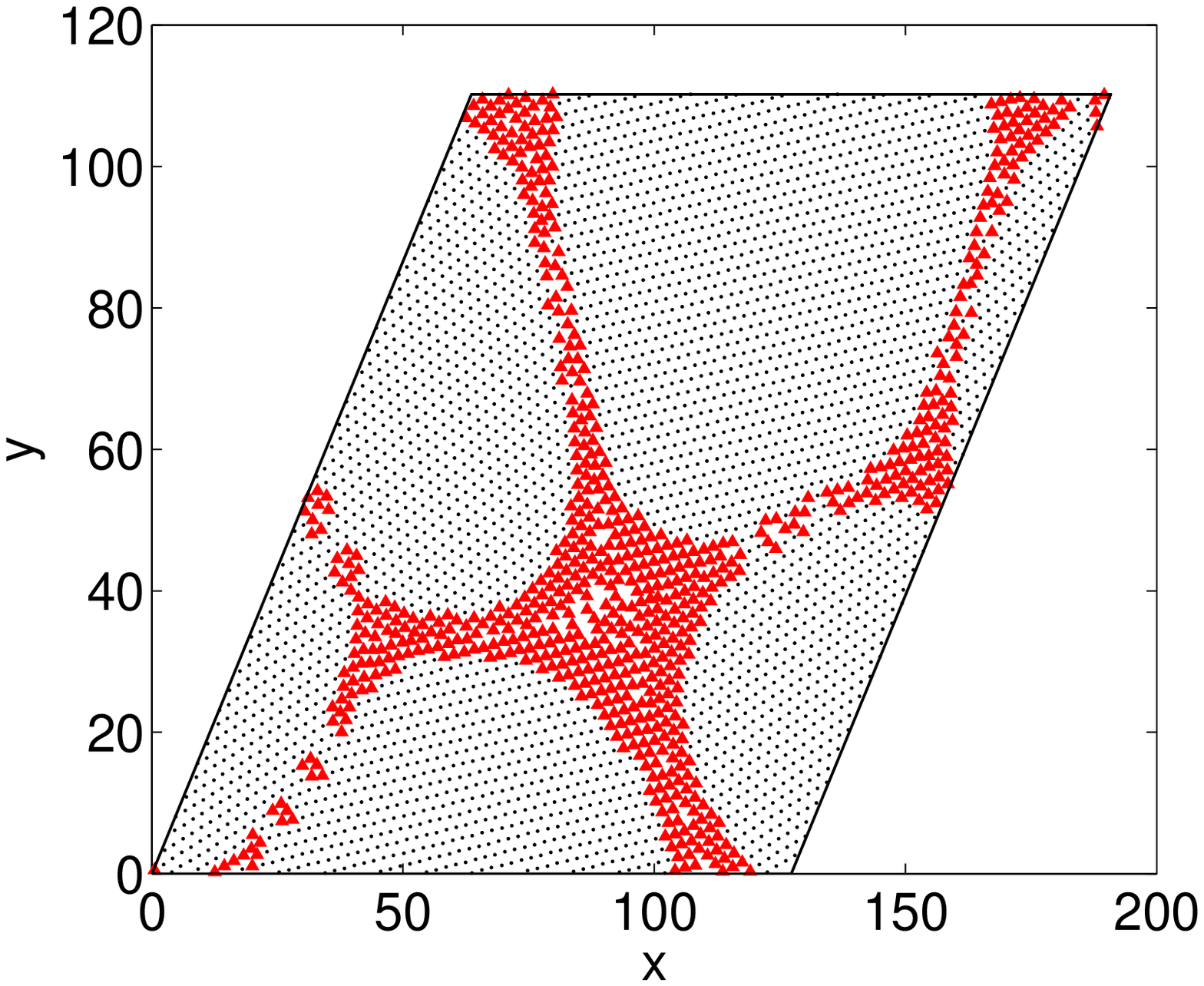}
\caption{(Color online) 
A metastable polycrystalline vortex configuration obtained as 
explained in the text. The temperature of the vortex system is $T=18.2$ K, slightly
below the melting temperature of the crystalline solid.  Distances are
in units of $a_0$ (see text). %otvf
The top panel shows the %otv
results of a Voronoi construction (see text) %otv
that brings out the details of the polycrystalline
structure. In this plot, vortices with 4, 5, 6 and 7 neighbors are shown as 
(green) triangles, (red) circles, (black)
dots and (blue) squares, respectively. Adjacent pairs of 5- and 7-coordinated sites 
correspond to dislocations which
line up along the grain boundaries that separate the microcrystals. The bottom panel shows the 
degree of localization of the vortices in the same configuration. Here, (red) triangles and 
(black) dots respectively represent liquid-like (less localized) and solid-like (strongly localized)
vortices. This plot illustrates the phenomenon of premelting along the grain
boundaries.}
\label{fig1}
\end{figure}

%otvr some editing in this paragraph, mostly things commented out
An example of such polycrystalline configurations, obtained by warming
up the polycrystalline state originally obtaind at a vortex temperature %otv too mmany times?
$T=17.0K$ to $18.2K$, is shown
in Fig.~\ref{fig1}. There
we plot the average positions of the vortices, defined to be the
computational lattice sites at which the density has a local maximum. 
%CDN defined the unit of length
In this and subsequent figures, the unit of length % used for plotting the 
%positions of the vortices 
is $a_0$, defined by 
the relation $\pi a_0^2 n_v=1$
where $n_v$ %otvr said earlier, the number of vortices per unit area, is given by
is related to the magnetic induction $B$ 
(which
was $B=0.2 T$ in the case shown) by
$n_v=B/\Phi_0$ where $\Phi_0$ is the superconducting flux %otvf
quantum. In
the top panel, we have shown the results of performing a Voronoi construction
for the vortex configuration. 
%cdr A Voronoi construction is the same as the usual
%division of a regular lattice into Wigner-Seitz cells, but performed on
%an arbitrary lattice: 
The Voronoi cell associated with a lattice point is %otv
the region of space nearer to that point than to any other lattice
point and the number of sides of the Voronoi cell represents the number of
neighbors of the lattice point. In the top panel, vortices with six
neighbors are shown as (black) dots and those with five or seven neighbors are shown as 
(red) circles and (blue) squares respectively. %otv colors added
A neighboring pair of five- and seven-coordinated vortices
constitutes a dislocation and grain boundaries correspond to arrays of such dislocations.
%In the plot shown in the top panel, 
Microcrystals can be seen as ordered regions consisting of sites with six neighbors,
and these microcrystals are separated by arrays of dislocations
(pairs of sites with five and seven neighbors)
denoting the grain boundaries. 

\begin{figure*}[tbh]
\includegraphics[width=3in]{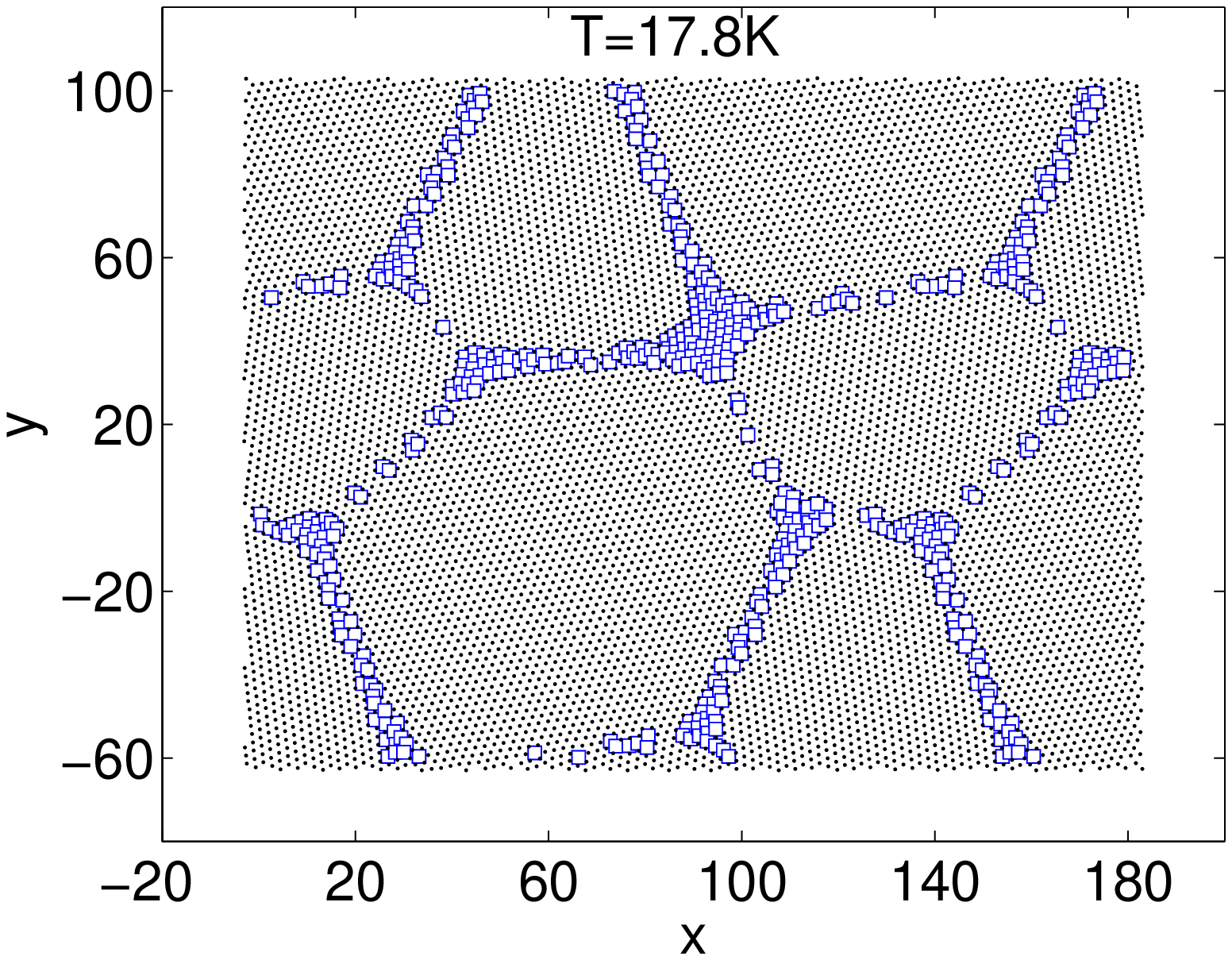}
\includegraphics[width=3in]{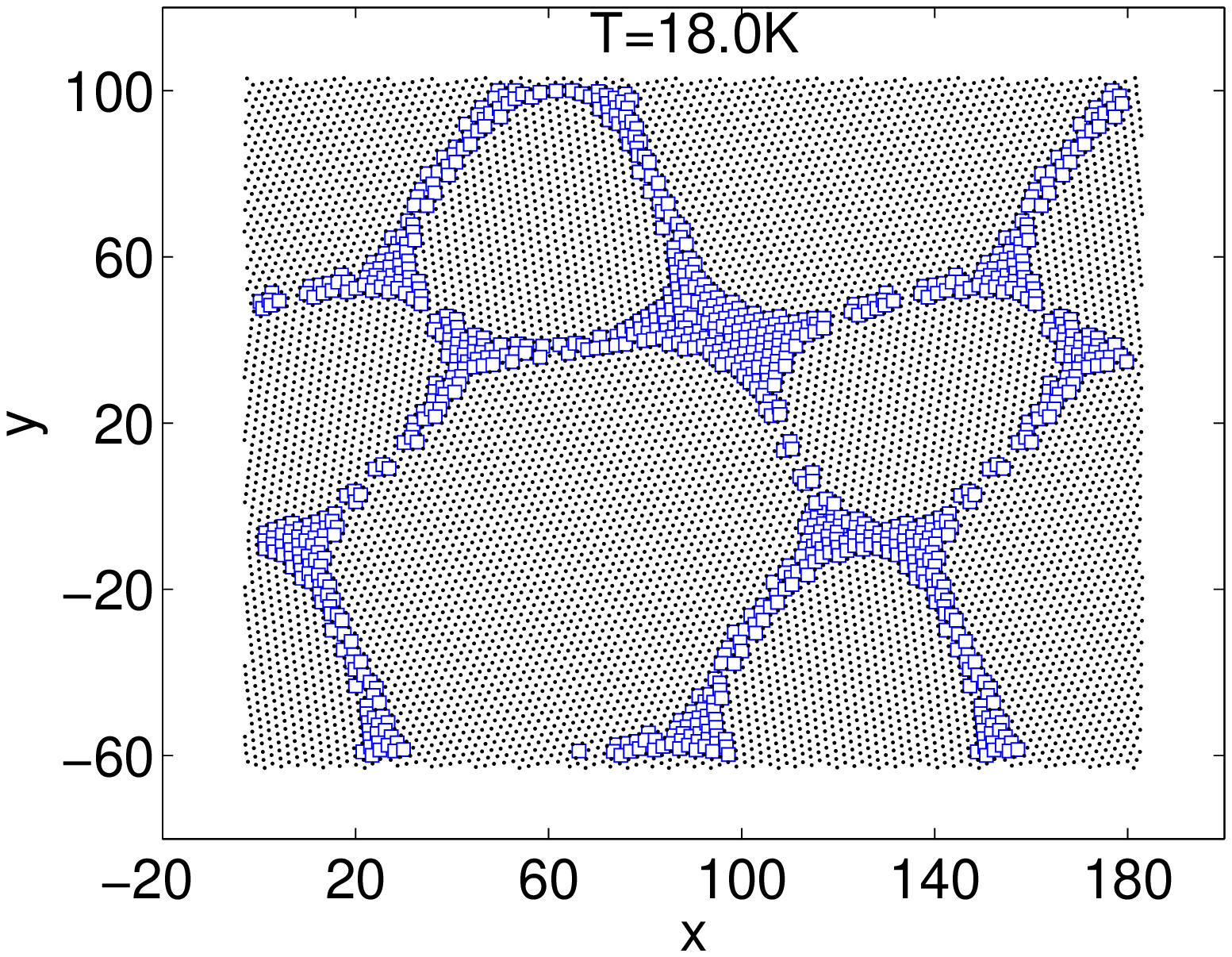}
\includegraphics[width=3in]{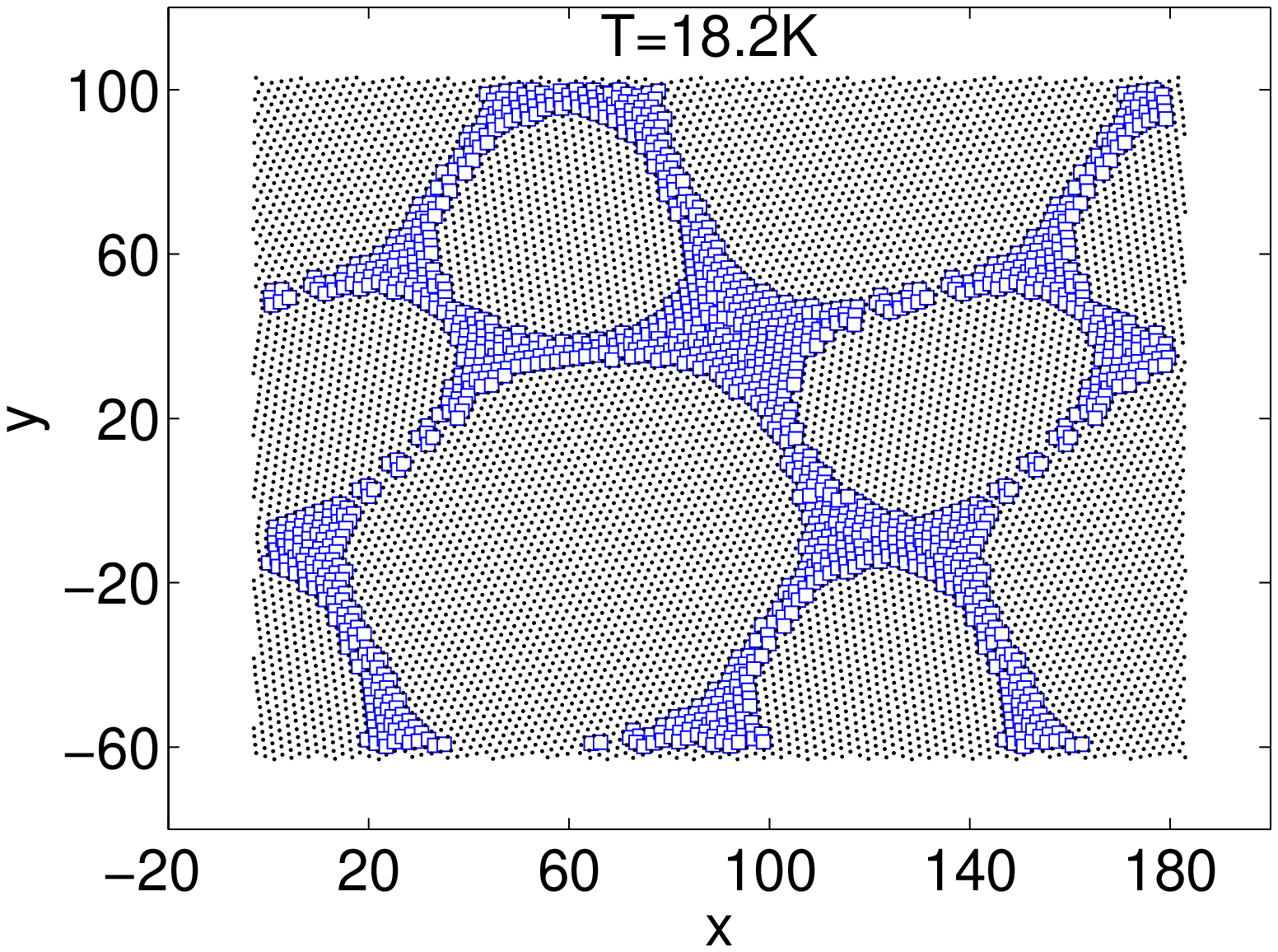}
\includegraphics[width=3in]{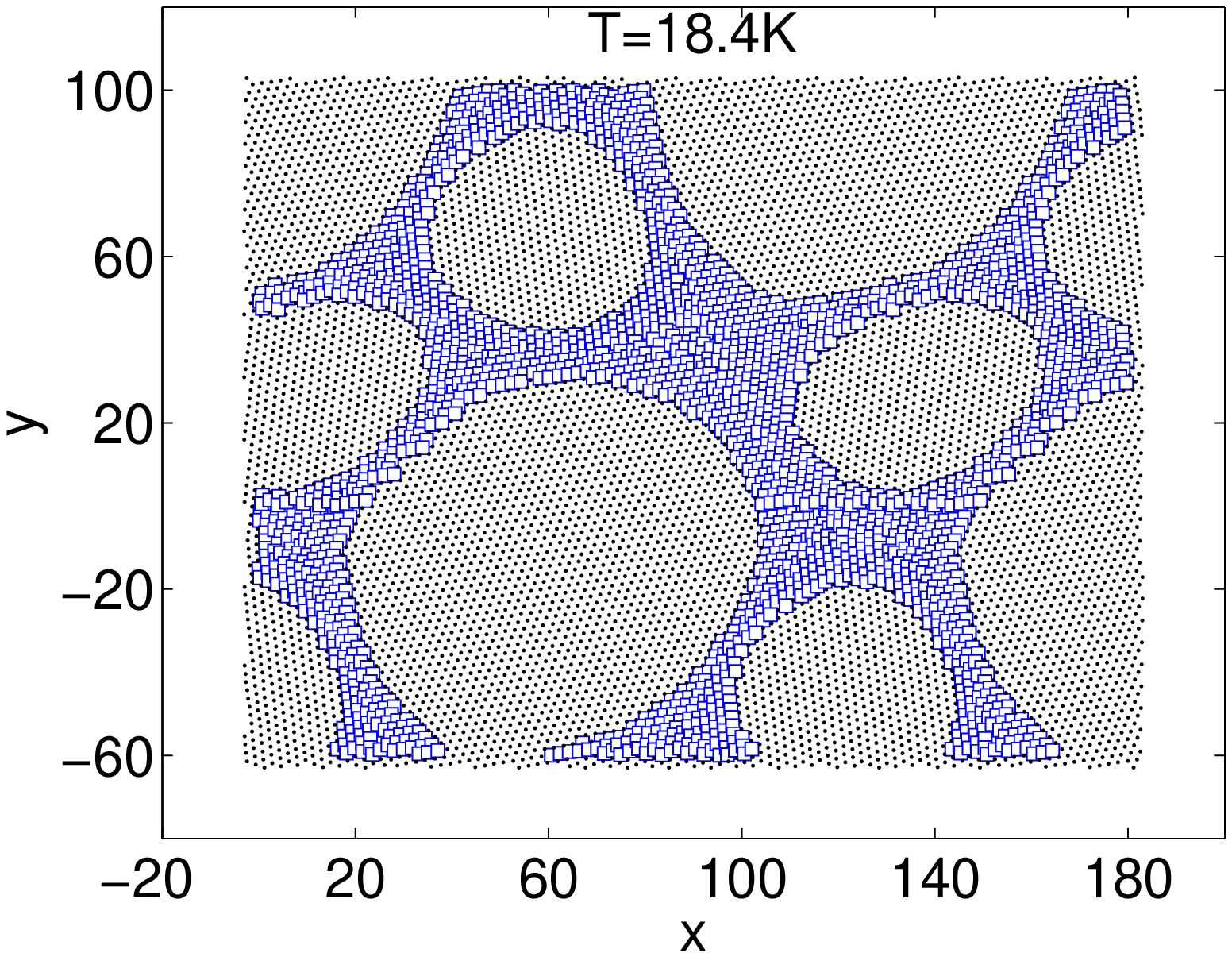}
\caption{ (Color online)
Evolution of the network of liquid channels with the temperature $T$ of the
vortex system in one of the samples studied. Each of the four
panels corresponds to a value of $T$ as indicated. Vortices in 
liquid regions  %otvf
are shown as (blue) squares and those in the microcrystal are shown as (black) dots.
One can see that as $T$ increases, the liquid channels become connected and 
closed loops of size comparable to that of the sample are formed.
Units of length are as in Fig.~\ref{fig1}.} %otvf
\label{fig2}
\end{figure*}

In the bottom panel of Fig.~\ref{fig1}, we show, for the same configuration, %as 
%the one in the top panel, 
the spatial distribution
of the value of the vortex density at each local density
%CDN changed rho_0 to n_v which was defined earlier as the vortex density
peak, normalized to the average vortex density value $n_v$. 
%otvr I did not touch this part, as additions have to be made!
The value of this dimensionless quantity provides a measure of the degree of
localization of the corresponding vortex.
%cdrev added a few lines to explain how we determine whether the vortices are liquid
%or solid-like.
Using the value of this quantity, we can determine whether the vortices in a
given region are liquid- or solid-like.
In our earlier studies of the vortex system~\cite{dv1,dv2,dv3} (see, for example,
Sec. IIIC of Ref.~\onlinecite{dv3}, or Sec. IIID of Ref.~\onlinecite{dv2}), %otv
we found that vortices in liquid regions correspond to local density peaks for which
this quantity has values of three or less, 
while those with higher values of this quantity correspond to solid
regions. In the plot, local density peaks representing solid and liquid %otv
regions according to this quantitative criterion are denoted as (black) dots and (red) triangles,
respectively. 
%otvr here I start editing again 
One can  see that all the vortices in the ordered regions are
solid-like: these regions correspond to microcrystals. In contrast, the
vortices near the grain boundaries form liquid
channels that separate the microcrystals. 
%cdrev end of changes
This plot illustrates the
occurrence of a kind of ``premelting'' in the regions near the grain boundaries: these
regions melt and become liquid-like (superfluid in the equivalent bosonic system) at temperatures
lower than that at which the bulk crystal melts (which we recall is $18.4K$).
%otvr , slightly higher than the
%otvr value, $18.2K$, for which the results shown in this figure were obtained). %otv
Premelting 
along grain boundaries is well-known~\cite{gbmelt} in classical solids. Our
observation of this phenomenon in the vortex system is consistent 
with experimental results. %otv paragraph below merged, sentence deleted
%cdr In this plot, the liquid channels that separate the microcrystals 
%do not yet percolate through the sample -- percolation of liquid regions occurs at a slightly
%higher temperature. 
At lower temperatures, the liquid channels along the grain boundaries are
narrower and obstructed at the narrowest spots (see Fig.~\ref{fig2} below for an example), forming
a chain-like structure of small liquid droplets. As the temperature
is reduced further, the liquid regions near the grain boundaries disappear completely and are
replaced by narrow strips of disordered solid.

%otvr paragraph below left alone
These results imply, from the vortex-boson mapping, that a two-dimensional polycrystalline
system of bosons exhibits superfluidity along grain boundaries over a suitable range of
parameter values. Using the realistic networks of grain boundaries obtaind from our numerical
studies of the vortex system, we can then investigate how the NCRI arising from the presence of 
superfluid channels in the bosonic system depends on the system parameters.
The numerically obtained samples in themselves are too small to allow
a realistic study of the flows, but a sufficiently,
indeed arbitrarily large, sample can be obtained from them through periodic repetitions (periodic
boundary conditions are used in our original numerical studies) that produce a
tiling of a larger region with the smaller samples. One obtains then a larger sample
for which the short-distance structure is the same, but which
has a different large-scale structure because longer liquid channels
are produced when different copies of the original sample are juxtaposed. Indeed, the
length of the largest channels after such a juxtaposition is of the order
of that of the combined sample. It is true that the structure of the original
small sample is preserved in the small-scale structure of the large
sample, which has an artificial periodicity. However, this is not important:
the behavior of the moment of inertia of the liquid region, and therefore the
NCRI is of course determined essentially by the largest-scale channels
in the problem. Scaling the system by a linear factor of $s>1$ in this way
introduces liquid channels that are larger by a factor of up to $s$ than the
original ones, while leaving the original small-scale structures and the channel
widths unchanged. It follows from geometrical and similarity considerations
that the overall moment of inertia of the channels, including the
contributions from the additional smaller structures, should then scale as
$s^4$. Hence, our conclusions for the moments of inertia, which we will
normalize to the overall rigid-body value, which also scales in the
same way, will be independent of this procedure. 
The samples considered in our NCRI calculations are selected portions of $3\times3$ tilings
of the originally obtained samples. 

%otvr light edits in this parag
The evolution of the network of liquid channels in one of these samples with 
increasing temperature $T$ of the vortex system is
displayed in  Fig.~\ref{fig2}. Each panel 
shows a different value of $T$ ranging from $17.8K$ to $18.4 K$. 
%cdr As in the bottom
%panel of Fig.~\ref{fig1}, 
In these plots, vortices in the liquid
regions %(which, to repeat, are identified with the regions in which the 
%local peak densities do not exceed $3n_v$) 
are shown as (blue) squares,  and those in the solid regions  as
(black) dots. At the lowest temperature, $T=17.8 K$, the liquid regions
form largely disconnected, small beads located along the grain boundaries, with
larger droplets forming at the intersections of two grain boundaries.
As $T$ is increased, the bead-like
liquid structures coalesce into channels, so that at about $T=18.2 K$,
connected structures of size comparable to that of the whole sample 
(in particular, a  ring-like structure that can be seen in the upper left corner of
the figure) appear.  Finally, at $T=18.4$ K, just below
the melting temperature of the vortex solid, the liquid regions have
fully developed into open channels which percolate through the sample.
As the superfluid in the equivalent bosonic system is
allowed, in the configurations obtained in the higher $T$ range, to freely flow along these
channels, whose characteristic size is now of the order of the sample 
size, it is clear that the moment of inertia will exhibit a substantial
reduction, leading to appreciable values of the NCRIF.
 
To numerically calculate the moment of inertia of the bosonic samples, %otv
we proceed as follows: we imagine the microcrystalline
sample rotating about its center of mass, with angular speed $\Omega$.
The solid regions of the sample, composed of the microcrystals,
rotate of course as a rigid body with the angular velocity
${\bf \Omega}$. The material in the liquid (superfluid in the bosonic system) channels
will flow according to a pattern that can be determined by
numerically solving the fluid flow equations of an
incompressible, irrotational (super)fluid.~\cite{fetter,hydro} These
equations can be solved to
obtain the velocity field ${\bf v}({\bf r})$ in the fluid regions.
The velocity field at the boundaries of the fluid part
of the sample must satisfy
the boundary condition\cite{fetter} that follows from assuming
that the fluid is confined by the rigid walls. This implies that the component
of the  velocity field ${\bf v}({\bf r})$
along the outward normal $\hat{\bf n}$ 
at any point on the boundary must be equal to the component 
of the rigid-body velocity ${\bf \Omega} \times {\bf r}$ 
along $\hat{\bf n}$ at that point:
\begin{equation}
{\bf v}({\bf r})\cdot \hat{\bf n} =({\bf \Omega} \times {\bf r})\cdot \hat{\bf n}
\label{bc}
\end{equation}
where ${\bf r}$ is a vector from the center of rotation to a point on the 
liquid region boundary. 
%CDN inserted zero circulation condition and new references
In addition, if the fluid region is not simply connected, then the circulation of the velocity
field along a closed path surrounding each inner boundary of the region must be specified in 
order to make the problem well-defined~\cite{mehlzimm}. We assume that the sample is
accelerated from rest at zero temperature, so that the values of these circulations
are zero~\cite{fetter67}. Once the velocity field is obtained, the 
resulting angular momentum, and hence the moment of inertia can be straightforwardly obtained
by numerical integration.

Analytic solution of this problem is feasible\cite{fetter,hydro} only %otv
for reasonably simple geometries: for the very irregular %otv
geometries involving the multiple channels and regions under study %otv
here solution can only be undertaken numerically. 
%cdrev reference to numerical method
The numerical method we have used involves spatial discretization and iterative
relaxation~\cite{numerics} to obtain a solution of the Laplace equation (see below)
with appropriate boundary conditions.
There are certain technical %otv
difficulties involved in doing this.  The main one is that our definition of whether a 
part of the sample is solid or liquid is based on the values of the density at 
the local  peaks which form a discrete lattice. This lattice %otv
must therefore be our computational
lattice for the purpose of calculating the velocity field and
the moment of inertia and we cannot truly gain %otv
additional precision by adding more points. Any attempt to do so
would necessarily involve  some non-controllable interpolation scheme. This unavoidable
situation leads to some uncertainties in our numerical results. In particular, the
final results for the rotational inertia include %otv reworded
some numerical uncertainty arising from this minimum spacing of our computtaional
lattice.
%CDN deleted a line that is repeated later 
%This uncertainty is relatively small, however, and does not affect our
%conclusions. %otv reworded and paragraphs merged some things moved to below
%otv depend to some extent on the 
%otv definition of the discretized versions of spatial derivatives. 
Another technical difficulty is that the boundary condition for
the velocity field, Eq.(\ref{bc}), is of the Neumann form if one uses
the standard method~\cite{fetter,hydro} to express 
the incompressible velocity field ${\bf v}({\bf r})$ as 
\begin{equation}
v_x({\bf r}) =-\partial \Psi({\bf r}) /\partial y,\,\,v_y({\bf r}) 
=\partial \Psi({\bf r}) /\partial x,
\label{sf}
\end{equation}
in terms of a stream function $ \Psi({\bf r})$. %otv
The simplication of replacing
the boundary condition of Eq.~(\ref{bc}) by the Dirichlet boundary condition,
$ \Psi({\bf r}) = \frac{1}{2} \Omega r^2$, at all points on the boundary, which 
was used in earlier studies~\cite{fetter,hydro}, works for some simple %otv
geometries, but leads to incorrect results when unobstracted %otv
ring-like channels are
present. This can be easily checked by applying this method to determine the
velocity field in a superfluid confined between two concentric cylinders %otv
rotating about their common axis. This forces us to use the original
Neumann boundary  conditions  of Eq.~(\ref{bc}). These are more difficult and awkward to %otv
implement numerically, as they involve %otv
the outward normal $\hat{\bf n}$ at the boundary, which can be defined in
different ways when the boundary consists of a set of discrete points. 
With the possibility of reducing the computational lattice spacing not being %otv added
available to us, this introduces additional numerical uncertainty, as %otv 
the final
results are somewhat sensitive to the way in which $\hat{\bf n}$ is defined.
However, these uncertainties %otvf repeted later, which are $\sim$ 25\% in the worst cases, 
do not seriously affect the
quantitative conclusions that we can draw from our calculation,
as we shall see below. %otvf
We recall %otv added may be superfluous
that using a scalar potential for the velocity field leads to other
undesirable problems~\cite{hydro} even in some analytic cases.

%cdrev more on numerical method
The irrotational nature of superfluid flow in the absence of any superfluid vortex implies that
${\bf\nabla} \times {\bf v} =0$ in the superfluid regions. This condition and
the definition of the stream function $\Psi$ in Eq.(\ref{sf}) imply that
$\Psi$ must satisfy the Laplace equation,
$\nabla^2 \Psi({\bf r})=0$, inside each superfluid channel. 
Using a triangular computational grid of spacing
equal to the average inter-particle distance and representing the Laplace operator by symmetric
differences~\cite{numerics}, the Laplace equation can be reduced to a set of linear equations
satisfied by the values of $\Psi$ at the computational grid points in the interior of each
superfluid region. The boundary conditions
of Eq.(\ref{bc}) can also be written as a set of linear equations involving the values of
$\Psi$ at the grid points on the boundary of a superfluid region and their nearest neighbors.
This set of coupled linear equations is numerically solved using iterative relaxation~\cite{numerics}.
This yields the values of $\Psi$ at the grid points, from which the velocity field and the rotational
inertia can be easily obtained.

\begin{figure}[tbh]
\includegraphics [width=3.5in,angle=0]{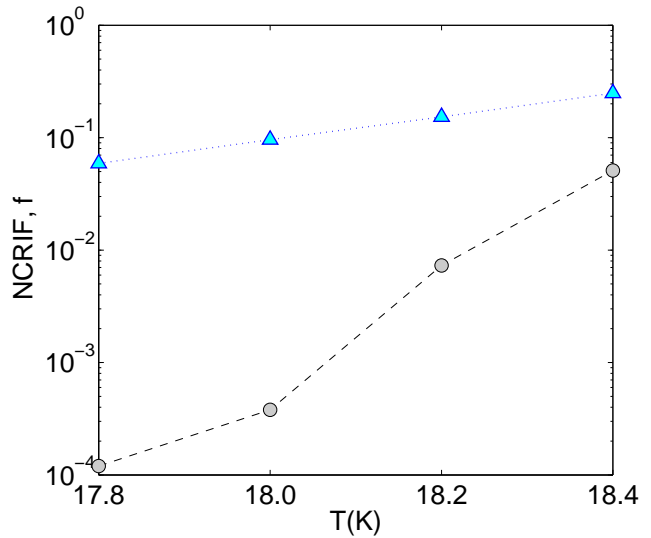}
\caption{The NCRIF, calculated and normalized as explained in the text, plotted 
as a function of $T$, the temperature of the vortex system (circles). 
%CDN only one set of data
%At each temperature, two
%data points (circles and triangles) are shown, representing the results obtained from two
%different ways of doing the numerical calculation of the velocity field. i
Also shown are 
the values of $f$, the liquid (superfluid in the bosonic system) fraction (triangles). The
straight line segments join consecutive data points.}
\label{fig3}
\end{figure}

%cdrev end of new text
In Fig.~\ref{fig3}, we present what is the main result of these
computations for the NCRI. The quantity plotted there (circles) as a function
of temperature $T$ (which, we reiterate, is the temperature at which the 
calculations for the vortex system were performed -- it should not be confused with the
temperature of the equivalent bosonic system which is  zero) %otv 
is the NCRIF, defined as the difference between the moment of inertia of
the entire sample as a rigid body and its actual moment of
inertia when the superfluid flow in the liquid portions is taken
into account, normalized by the total rigid-body moment of inertia. 
These results are for the sample shown in Fig.~\ref{fig2}. 
%CDN We now show only one set of data
%The
%two sets of data points, (triangles and circles) %otv
%for each value of $T$ represent the results obtained from two
%different ways of implementing the numerical calculation, %otv
%using different %otv definitions for the 
%discrete 
%forms of spatial derivatives and outward normals. The difference between the
%two sets of data points provides a measure of the numerical %otv 
%uncertainty in the %otv numerically
%obtained values of the NCRIF.
At lower temperatures, when the liquid regions form small, disconnected clusters (see the
discussion above in connection with Fig.~\ref{fig2}), the superfluid
flow is negligible and the entire system 
behaves as a rigid body, so the value of the NCRIF is very close 
to zero. At higher temperatures, when the disconnected liquid regions join together 
to form extended channels, superfluid
flow begins to occur over relatively large portions of the sample and
eventually the characteristic channel length becomes of the order
of the system size. At $T=18.2K$, a large, ring-like channel opens up 
(see Fig.~\ref{fig2}), providing a closed path for the flow of the superfluid.
This drastically reduces the moment of inertia;
recall for example that the moment of inertia of a superfluid confined in a ring that
is rotating about its center is zero. Therefore the NCRIF increases
very drastically (note the logarithmic vertical scale) and by the time
the temperature is reached where the channels are fully open, it has
increased by nearly three orders of magnitude, while of course
remaining relatively small ($\sim$ 5\%). Our conclusion about a %otv
rapid increase in the value of the NCRIF at $T=18.2K$, when a large, 
ring-like channel
opens up to allow the flow of the superfluid along a closed path, is not 
affected by
the above mentioned uncertainties in the numerical results, which  %otvf, as mentioned above, 
are $\sim$25\%
in the worst case.
%: both methods of calculation show a large
%increase in the NCRIF at this temperature. 
We have repeated the whole calculation for
a second polycrystalline configurations and obtained very similar results. Specifically,
we find in both cases %otvf is this right?
that the NCRIF increases by a factor of $\sim 18$ at $T=18.2K$ when a large,
ring-like liquid channel opens up.

In Fig.~\ref{fig3}, we have also shown (triangles) %otv
the dependence of $f$, the fraction of the sample
that is liquid-like (superfluid in the bosonic system), on $T$. It is clear from this
plot that the dependence of this quantity on $T$ is qualitatively different from that of
the NCRIF: the liquid fraction $f$ increases smoothly with increasing $T$ and does not
exhibit the rapid increase seen in the NCRIF at $T=18.2K$. This observation brings out
the important point that the NCRIF may not provide a good measure of the fraction
of superfluid regions of the system if the superfluidity occurs along a network of
channels -- the connectivity of the channels plays an important role in determining the
value of the NCRIF. The NCRIF remains small as long as the network does not contain
large ring-like paths along which the superfluid can flow without any blockage. The 
first opening up of such paths as some system parameter is varied leads to a large increase 
in the value of the NCRIF. This large increase, which may look like the onset of 
superfluidity, does not necessarily correspond to a sudden, large increase in the 
superfluid fraction $f$ of the system. This should be kept in mind while interpreting
experimental data for the NCRIF.

\section{Summary and discussions}
\label{summary}

In this work, we have studied the NCRI  
arising from superfluidity along grain boundaries %otv
in a two-dimensional bosonic system
at zero temperature.
By making use of standard mappings, we have related the properties
of the bosonic system to those of a system of superconducting
vortex lines at finite temperatures. The latter system has been extensively
studied by numerical methods and computationally
obtained structures in the different phases it can
exhibit are available. One of the possible structures of this vortex
system is a polycrystalline solid in which liquid regions near the  grain boundaries 
between microcrystals expand and form
more connected regions which eventually extend through the sample 
as the vortex system temperature is increased. These
vortex configurations, obtained from previous numerical studies, 
correspond, via the above mentioned mappings, to different configurations of the
boson system, each containing more or less extensively connected narrow 
superfluid regions, separating much larger 
and compact crystalline regions. We have then calculated the moment
of inertia of the bosonic samples by numerically solving the equations of
superfluid hydrodynamics in the superfluid regions of the sample.

We have found that the non-classical portion of the moment of inertia
of such samples (the NCRIF) is nearly zero as long as the vortex liquid (bosonic superfluid)
regions remain disconnected, but it increases abruptly by about two orders
of magnitude as the superfluid channels become connected to form ring-like structures 
with sizes comparable to the system size. The  values we find
for the fully developed NCRIF are of the order of a few percent. Thus, our results indicate that
NCRI behavior can indeed arise in two-dimensional bosonic systems from superfluid
regions associated with grain boundaries. 
The magnitude of the jump in the NCRIF and the final NCRIF values obtained from
our calculations are quite similar to what is found experimentally in solid $^4{\rm He}$. Our results,
therefore, lend support to the notion that supersolid phenomena in
$^4{\rm He}$ are related to superfluidity in grain boundaries
or other crystal defect regions, such as dislocations. Our results also indicate that an abrupt
increase in the NCRIF from a vanishingly small value to a few percent does not necessarily
correspond to a similar increase in the fraction of the sample that is superfluid. 
As shown in Fig.~\ref{fig3}, such an increase in the NCRIF may actually correspond to the opening up
of system-spanning ring-like channels through which the superfluid can flow without any 
block. The importance of the existence of a percolating network of superfluid channels 
in observing macroscopic signatures of superfluidity has been pointed out in
an earlier study~\cite{toner2}. However, we are not aware of any calculation of the 
behavior of the NCRIF across the percolation transition.

As discussed above, the results of existing experimental~\cite{expt1,expt2,expt3} and 
numerical~\cite{dv1,dv2,dv3,sim1,sim2} studies of vortex lines with dilute columnar pinning 
strongly suggest, via the maping used here, %otv
that a polycrystalline state with
superfluidity along grain boundaries occurs in a two-dimensional bosonic system with a small
concentration of strong pinning centers. Such a state may be seen in experments~\cite{crowell} on
films of $^4{\rm He}$ adsorbed on substrates with imperfections that act as pinning
centers. In the vortex system, the equilibrium phase changes from
polycrystalline Bose glass to vortex nanoliquid and then %otv
to homogeneous liquid as the temperature is
increased. In the equivalent bosonic system, this sequence corresponds to going from a 
polycrystalline solid to a ``supersolid'' with superfluidity in the grain boundary regions
and then %otv
to a homogeneous superfluid. It is, however, not clear how the experimentally accessible
parameters in the bosonic system
can be changed to simulate the effects of increasing the temperature in the vortex system.
The same sequence of phases can also be seen~\cite{expt1,expt2,expt3} in the vortex system by
increasing the magnetic induction at a fixed, low temperature. Since the magnetic induction
in the vortex system determines the areal density of the vortex lines, the supersolid
phase in a two-dimensional bosonic system may %otv possibly 
be accessed by changing the coverage of the
$^4{\rm He}$ film. The supersolid would be the true equilibrium phase 
of the system %otv only  (saying only is contradicted by next sentence)
when random defects are present. %otv 
Our numerical results, furthermore, suggest that the supersolid %otv may
survives as a metastable phase even in the absence of 
random pinning. While it is not obvious %otv
how such a metastable phase %otv(if it exists) 
could %otv
be accessed in experiments,  the %otv
supersolid behavior observed in three-dimensional solid $^4{\rm He}$ must also be a metastable
phenomenon if it arises from superfluidity in a network of structural defects, because these
defects would not be present in the true equilibrium solid which should be a perfect crystal.

%cdrev new text
In making any comparison of the behavior of the bosonic system found in our study with
experiments on $^4{\rm He}$ samples, one should keep in mind the important fact that the interaction
between two $^4{\rm He}$ atoms (strongly repulsive at short distances, with a weak attractive
part at larger distances that falls off with distance as a power-law) is substantially different 
from that between two bosonic
particles in the system considered here (purely repulsive, logarithmic at short distances and
exponentially decaying at longer distances). For this reason, the results of our study can not
be applied quantitatively to experimentally studied $^4{\rm He}$ systems. However, we expect 
the qualitative behavior found in our study to be observed in systems of $^4{\rm He}$ atoms because
the phenomena on which our conclusions are based (i.e. the formation of polycrystalline 
structures and ``pre-melting'' along grain
boundaries) are fairly generic, independent of the details of the inter-particle potential.
%cdrev end of new text

%cdrev new lines
In experiments on $^4{\rm He}$, a normal solid to supersolid transition is observed
~\cite{chan1,chan2,balibar,reppy,kojima,kubota,kondo,hunt,beamish} as the temperature of the 
sample is decreased. So, it is interesting to inquire whether our study provides any 
information about the behavior of the bosonic system as {\it its} temperature is changed.
Unfortunately, we can not address this experimentally relevant question in our study. This 
is because a nonzero temperature in the bosonic problem maps to a vortex system
of finite thickness $L$, and our studies of the vortex system, carried out
for the $L \to \infty$ limit, do not provide any information about phase
changes that may occur as $L$ is varied at constant temperature.
% end of new text

We close with a discussion of whether the available experimental results for supersolid
$^4{\rm He}$ show any evidence for the sequence of phases (polycrystal to vortex nanoliquid to
vortex liquid) found in the vortex system upon increasing its %otv
temperature $T$. %otv We recall that
In the equivalent bosonic system, this sequence correspond to a transition from a defected solid
to a supersolid, followed by a second transition from the supersolid to a superfluid. %otvAlso, 
The %otv it should be mentioned on page 8 that we discuss this later (here).
vortex-boson mapping implies that increasing $T$ in the vortex problem is 
equivalent to increasing the relative importance of quantum effects in the zero-temperature
boson problem. It has been suggested~\cite{nosanow,chan3} that increasing the pressure $P$ has the
effect of reducing the relative importance of quantum fluctuations in solid $^4{\rm He}$. 
Assuming this to be correct, increasing (decreasing) $T$ in the vortex system should be analogous to
{\it decreasing} ({\it increasing}) $P$ in  $^4{\rm He}$ experiments. Then 
the relevant question is whether a sequence of normal solid to supersolid to superfluid transitions
occurs in  $^4{\rm He}$ as $P$ is decreased at constant temperature.
There is some experimental evidence~\cite{chan3} for a reduction in the low-temperature value of
of the apparent superfluid fraction in supersolid  $^4{\rm He}$ as $P$ is increased beyond
55 bar. This may correspond to a transition from the supersolid to a regular defected solid,
analogous to the disappearance of the liquid regions in the vortex system as $T$ is decreased.
Also, the ``phase diagram'' of $^4{\rm He}$ in the $P-T$ plane shown in Ref.~\onlinecite{chan1}
suggests that the temperature of the normal solid to supersolid transition decreases
as $P$ is increased (the transition temperature is found to decrease from 315 mK at a 
pressure of 26 bars to 230 mK at pressures exceeding 40 bars). If this 
relatively weak dependence of the transition temperature on pressure is a genuine effect, 
then decreasing $P$ at constant $T$ would indeed lead to
a sequence of two transitions, first from the normal solid to the supersolid and then from
the supersolid to the superfluid phase. This would be analogous to the behavior seen in our 
calculations. More detailed investigations of the pressure-dependence of supersolid behavior in
$^4{\rm He}$ would  be very interesting in this context.

% The ruledtabular environment adds doubled rules to table and sets a
% reasonable default table settings.
% Use the table* environment to get a full-width table in two-column
% Add \usepackage{longtable} and the longtable (or longtable*}
% environment for nicely formatted long tables. Or use the the [H]
% placement option to break a long table (with less control than 
% \begin{table}%[H] add [H] placement to break table across pages
% \begin{ruledtabular}
% Lines of table here ending with \\
% \end{tabular}
% \end{ruledtabular}
% \end{table}

%\appendix
%\section{}

\begin{acknowledgments}
This work was supported  in part by NSF (OISE-0352598) and by
DST (India).
\end{acknowledgments}


\end{document}